\documentclass{revtex4}  
\usepackage[latin1]{inputenc}
\usepackage{amsmath}
\usepackage{amsfonts}
\usepackage{amssymb}
\usepackage{graphicx,color}

\begin{document}

\title{Computational temporal ghost imaging}
\author{Fabrice Devaux$^{1*}$, Paul-Antoine Moreau$^2$, S\'everine Denis$^1$ and Eric Lantz$^1$}
\affiliation{$^1$ Institut FEMTO-ST, D\'epartement d'Optique P. M. Duffieux, UMR 6174 CNRS \\ Universit\'e Bourgogne Franche-Comt\'e, 15b Avenue des Montboucons, 25030 Besan\c{c}on - France\\
$^2$ Present address : Centre for Quantum Photonics, H. H. Wills Physics Laboratory and Department of Electrical and Electronic Engineering, University of Bristol, Merchant Venturers Building, Woodland Road, Bristol BS8 1UB, United Kingdom}

\date{\today}
\email{ $^*$ fabrice.devaux@univ-fcomte.fr}

\begin{abstract}
Ghost imaging is a fascinating process, where light interacting with an object is recorded without resolution, but the shape of the object is nevertheless retrieved, thanks to quantum or classical correlations of this interacting light with either a computed or detected random signal. Recently, ghost imaging has been extended to a time object, by using several thousands copies of this periodic object. Here, we present a very simple device, inspired by computational ghost imaging, that allows the retrieval of a single  non-reproducible, periodic or non-periodic, temporal signal. The reconstruction is performed by a single shot, spatially multiplexed, measurement of the spatial intensity correlations between  computer-generated random images and the images, modulated by a temporal signal, recorded and summed on a chip CMOS camera used with no temporal resolution. Our device allows the reconstruction of either a single temporal signal with monochrome images or wavelength-multiplexed signals with color images. 
\end{abstract}

\maketitle
\section{Introduction}
Exploitation of the statistical properties of classical or non classical light sources is the cause of fascinating new applications. For the two last decades, ghost imaging has emerged as a way to form images of an object with a Single Point Detector (SPD) that does not have spatial resolution. The initial works used the quantum nature of entanglement of a two-photons state, where photons of a pair are spatially correlated, to detect temporal coincidences  \cite{pittman_optical_1995}. While one of the photons passing through the object was detected by a photon counter with no spatial resolution, its twin photon was detected with spatial resolution by scanning the transverse plane with a single detector\cite{pittman_optical_1995}, or recently by an intensified charge-coupled device (ICCD) \cite{morris_imaging_2015}.

Later, ghost imaging exploiting the temporal correlations of the intensity fluctuations of classical \cite{bennink_two-photon_2002} or pseudothermal light \cite{ferri_high-resolution_2005} was proposed.
More recently, computational ghost imaging and ghost diffraction were performed with  only one SPD \cite{shapiro_computational_2008,bromberg_ghost_2009}: the object was illuminated by a pseudothermal light beam, generated with a Spatial Light Modulator (SLM) addressed with random phase masks. Then, the transmitted light was detected with the SPD. The image or the Fourier transform of the object was reconstructed by correlating the temporal fluctuations of the calculated field patterns with the measured intensities. With the same principle, ghost imaging with wavelength-multiplexing has been performed \cite{zhang_wavelength-multiplexing_2015}. At last, application of quantum ghost imaging to long distance optical information encryption and transmission has been demonstrated \cite{dong_long-distance_2016}.

By taking into account space-time duality in optics, the extension of the results of spatial ghost imaging to the time domain has been investigated theoretically, numerically and recently experimentally either with a classical non-stationary light source \cite{shirai_temporal_2010,setala_fractional_2010}, bi-photon states \cite{cho_temporal_2012}, a chaotic laser \cite{chen_temporal_2013} or a multimode laser source \cite{ryczkowski_ghost_2016}. In all cases, the light emitted by the sources was split into two arms, called "reference" and "test" arms. While in the test arm the light was transmitted through a "time object" and detected with a slow SPD which can not properly resolve the time object, in the reference arm the light, that did not interact with the temporal object, was detected with a fast SPD. As for spatial ghost imaging, the temporal object was reconstructed by measuring the correlations of the temporal intensity fluctuations or the temporal coincidences between the two arms. In \cite{ryczkowski_ghost_2016}, measurements over several thousands copies of the temporal signal were necessary to retrieve an embedded binary signal with a good signal-to-noise ratio \cite{faccio_optical_2016}.  

The extension of spatial ghost imaging to the time domain looks attractive for dynamic imaging of ultra-fast waveforms with high resolution. However, the currently proposed solutions require many realizations of the same temporal signal, limiting the current applications to the detection of synchronized and reproducible signals \cite{faccio_optical_2016}. This is in contrast with spatial ghost imaging, where the object is unique, but multiplied in the time domain by a random modulation, different from one pixel to another, leading to multiplexing in this time domain. In the present paper we propose an original and completely different scheme which is the exact space-time transposition of computational ghost imaging \cite{bromberg_ghost_2009} and of wavelength-multiplexing ghost imaging \cite{zhang_wavelength-multiplexing_2015}: a single shot acquisition of a non-reproducible time object is performed by multiplying it with computer-generated random images, ensuring spatial multiplexing of temporal intensity correlations before detection of the time integrated images with a camera that has no temporal resolution. Experiments are performed with monochrome images or color images to reconstruct either a single temporal signal or wavelength-multiplexed temporal signals, respectively.  

\section{Results}
\begin{figure}[htbp]
\centering
\includegraphics[width=8cm]{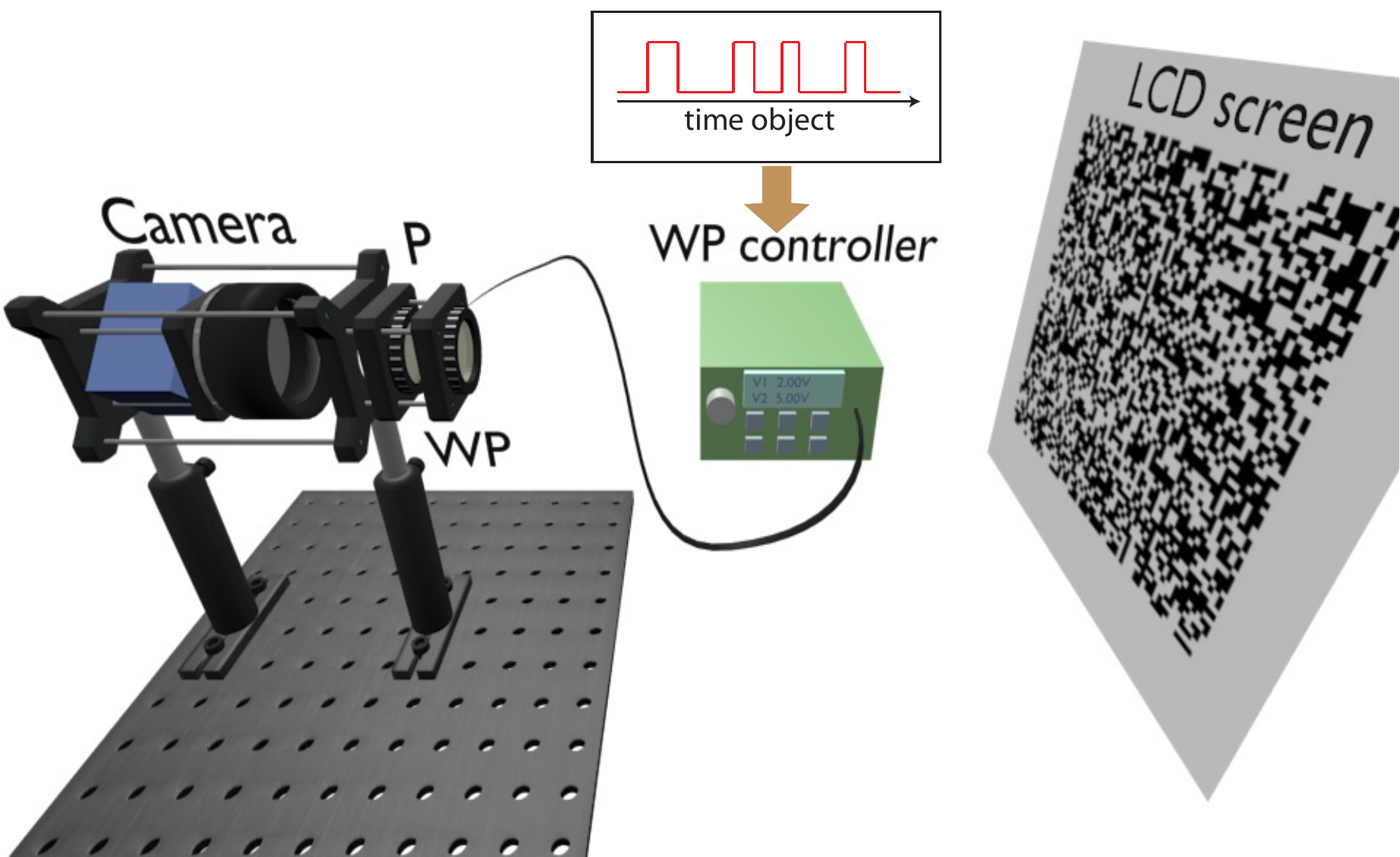}
\caption{ Experimental setup : (P) linear polarizer (WP) liquid crystal variable wave plate. Images of random patterns displayed on a LCD screen are acquired with a CMOS camera. The transmission of the imaging system is modulated by a temporal object that drives the WP placed in front of the polarizer.}
\label{setup}
\end{figure}
  
First, Let us consider the experiment with monochrome images. Fig. \ref{setup} depicts the set-up. A stack of $K$ independent random binary patterns $X$, with a large number of pixels, are generated and displayed on a LCD (Liquid Crystal Display) screen. In these binary patterns, a pixel takes the value 1 with probability  $p$ and the value 0 with  probability  $1-p$, hence verifies the statistical properties of a Bernouilli distribution ($\langle X\rangle  =p$, $\sigma^2_X=p(1-p)$). Images of the patterns are acquired with a compact CMOS USB2.0 camera (IDS UI-1640C, 1280$\times$1024 pixels) on a 8 bits grey scale. Since the light emitted by the LCD screen is linearly polarised at $45^\circ$, the energy transmission of the imaging system can be linearly modulated by a temporal signal during the exposure time of the camera, by means of a liquid crystal variable wave plate (WP, Thorlabs LCC1113-A) placed before a linear polariser (P) in front of the camera objective. 

At first, each pattern $X_k$ of the stack ($k$ denotes the realisation number of the pattern in the stack) is recorded individually with the same exposure time. The retardation of the WP and the direction of the polariser are adjusted to maximise the energy transmission of the imaging system, that will be considered in the following as a level of 100$\%$ ($T=1$). Then, the mean and the variance of each image $X^\prime_k$ are measured. Since the displayed patterns are binary, the sharp edges of the original pixels are blurred in the recorded images because of the modulation transfer function (MTF) of the objective. Consequently, as the recorded images are encoded over 255 grey levels, they do no longer verify the initial statistical properties. Several experiments were conducted to optimize all the setup parameters (focus, numerical aperture and magnification of the objective, number of independent pixels in patterns, probability $p$) such as the statistical properties of intensity fluctuations in the recorded images are the closest of those of the computed patterns (see supplement).

In a second experiment, the  patterns of the stack are displayed successively on the LCD screen, imaged with a transmission coefficient given by the  driving temporal signal ($0\leq T(t)\leq 1$), and summed on the camera during a long exposure time (5 to 10 $s$). In that case, the recorded image $S$ corresponds to the time integrated image of the displayed patterns such as the level of a pixel $S_{ij}$ of coordinates $(i,j)$ is given by : 

\begin{equation}\label{eq1}
S_{ij}=\sum\limits_{k=1}^{K}T(k)X^\prime_{k,ij}
\end{equation} 
where $T(k)$ is the value of the transmission at the time when the $k^{th}$ pattern is displayed.

\begin{figure}[htbp]
\centering
\fbox{\includegraphics[width=8cm]{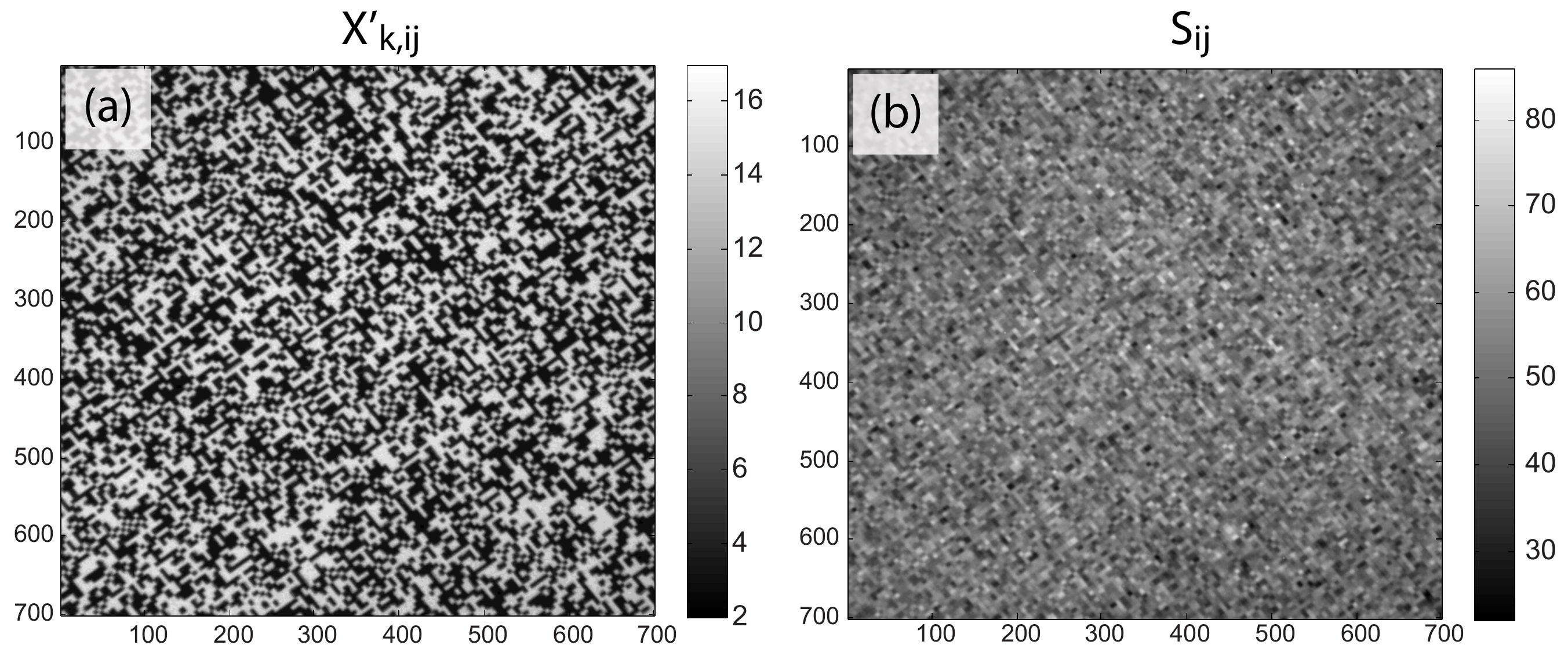}}
\caption{(a) Image of a single random pattern where $p=0.5$. (b) Image of the time integrated pattern's stack with a temporal object embedded. The measurements are performed inside a ROI of $700\times 700$ pixels.}
\label{XetS}
\end{figure}

Fig. \ref{XetS} shows a selected region of interest (ROI) in images of the displayed patterns. Fig. \ref{XetS}a corresponds to the image of a single pattern and Fig. \ref{XetS}b corresponds to the image of the time integrated patterns of the stack where a temporal object is embedded. Because the circular aperture of the WP limits the field of view of the imaging system and induces deterministic fluctuations for pixels close to the border, we selected the larger ROI of $N\times N$ pixels ($N=700$) in images where the gray level of a lighted pixel is almost constant. In the subsequent calculations, residual covariances between images of two independent binary patterns, due to these deterministic intensity fluctuations, are removed by filtering numerically the recorded images. Because of the magnification of the imaging system, the image size of an independent pixel of the displayed patterns is larger than the size of a pixel of the camera. Then, the number of effective independent pixels in an image is smaller than the number of pixels in the selected ROI.   

As for other methods of ghost imaging, the temporal signal is reconstructed by calculating the intensity correlations between the time integrated image $S$ and the images $X^\prime$ of the $K$ random patterns. The value of $T(k_0)$ at the "time" $k_0$ is estimated by (a hat means "estimator of"):
\begin{equation}\label{eq2}
\hat{T}(k_0)=\frac{\gamma\sum\limits^{N}_{i=1}\sum\limits^N_{j=1} \left(S_{ij}-\overline{S}\right)\left(X^\prime_{k_0,ij}-\overline{X}^\prime_{k_0}\right)}{\sum\limits^{N}_{i=1}\sum\limits^N_{j=1}\left(X^\prime_{k_0,ij}-\overline{X}^\prime_{k_0}\right)^2}
\end{equation}
where $\overline{S}$ and $\overline{X}^\prime_{k_0}$ are the arithmetic mean levels of the related images. $\gamma$ is a normalization coefficient corresponding to the ratio between the exposure time of the camera for the acquisition of the individual images $X^\prime_k$ and the display time of the random patterns during the acquisition of $S$. It can be shown (see supplement) that the noise in the measurement is minimized when $p=0.5$. Then, the signal-to-noise ratio ($SNR$) is given by:
\begin{equation}\label{eq3}
SNR[T(k_0)]=\frac{T(k_0)}{\sigma_{T(k_0)}}=\frac{N_{eff}T(k_0)}{\sqrt{\sum\limits^K_{k=1,k\neq k_0}T^2(k)}} \geq \frac{N_{eff}}{\sqrt{K-1}}T(k_0)
\end{equation}  
where $N^2_{eff}$ represents the total number of effective independent pixels in recorded images. With a ROI of $700\times 700$ pixels, $N^2_{eff}$ has been estimated as $153\times 153$ independent pixels (see supplement).

\begin{figure}[htbp]
\centering
\includegraphics[width=8cm]{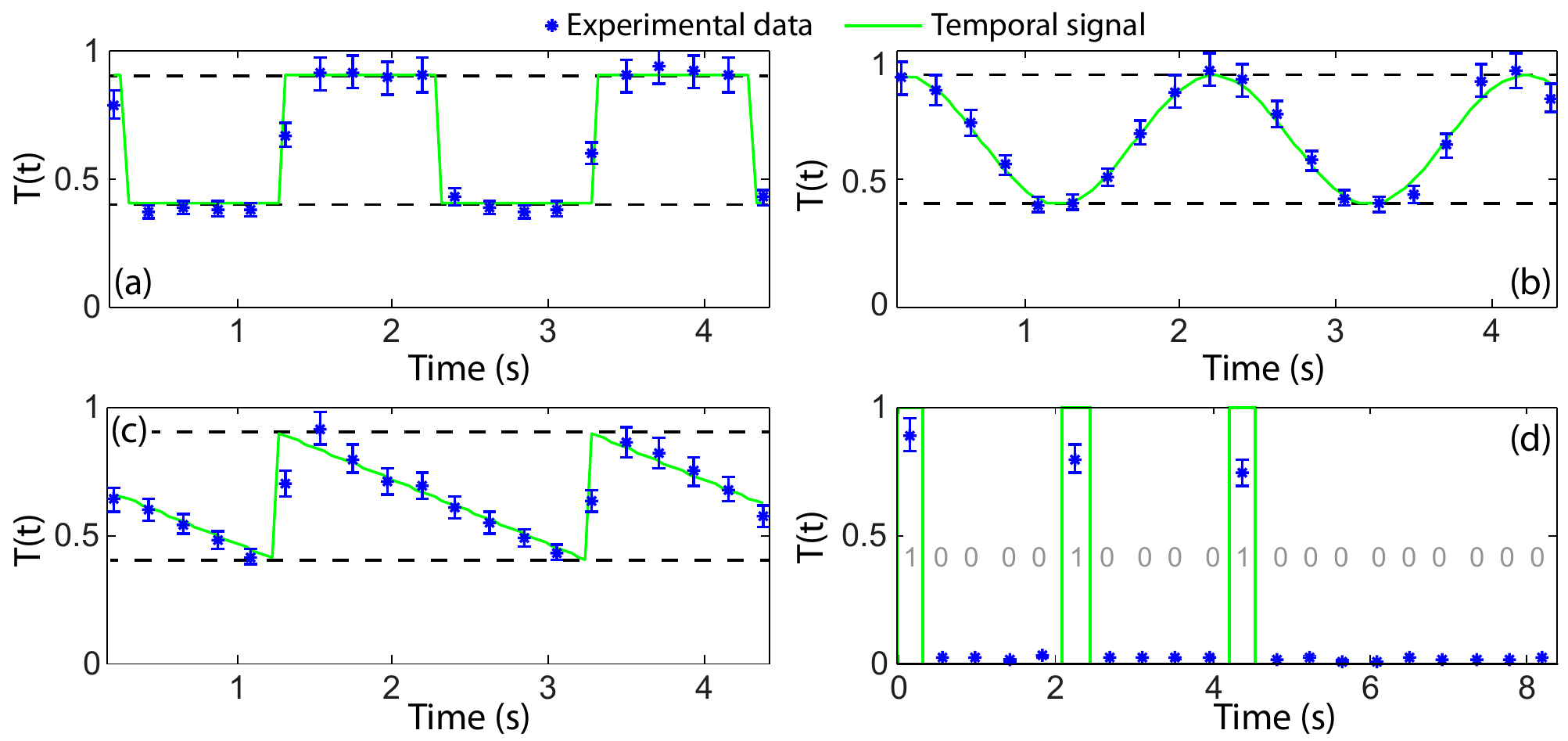}
\caption{Ghost images of different temporal signals. 3 kind of 0.5 $Hz$ periodic signals: (a) square, (b) sinus, (c) ramp. (d) corresponds to the single shot measurement of a binary word of 20 bits. Each acquisition is performed with the same stack of 20 random patterns. Blue stars correspond to experimental data and the green curves show the original temporal signals. Error bars are deduced from the measured $SNR$. For periodic signals, horizontal black dotted lines show the effective amplitude of the transmission levels of the imaging system.}
\label{signaux}
\end{figure}

Fig. \ref{signaux} shows ghost images of different temporal signals (analogic or binary and periodic or non periodic). Each ghost image corresponds to a single shot measurement which is performed with images of $700\times 700$ pixels  (where $N_{eff}=153$), the same stack of 20 random patterns (where $p=0.5$) and a long camera exposure time between 6 $s$ and 8 $s$. For periodic signals (figures \ref{signaux}a to Fig. \ref{signaux}c), the WP controller is driven in a linear regime by 0.5 $Hz$ periodic signals with different shapes (square, sinus and ramp) which are depicted by the green curves (only phases of these curves are adusted to fit the experimental data represented by the blue stars). Error bars are deduced from the measured $SNR$. Voltage amplitude of these signals are fixed such as the transmission of the imaging system is in the range $0.4\leq T(t)\leq 0.9$. These low and high transmission levels are depicted in Fig. \ref{signaux}a-c by the horizontal black dotted lines. Fig. \ref{signaux}d corresponds to the ghost image of a random binary word of 20 bits  formed with a mechanical shutter ensuring $T=0$ when closed, but $T<1$ when open because opening is not synchronized with the display of the random patterns on the LCD screen. These results clearly show that our device is able to reconstruct, with a single shot measurement and with a very good accuracy, various temporal analogic or binary signals. We emphasize that thanks to the single shot measurement, our device allows non reproducible and non synchronizable signals to be recorded.     

\begin{figure}[htbp]
\centering
\includegraphics[width=8cm]{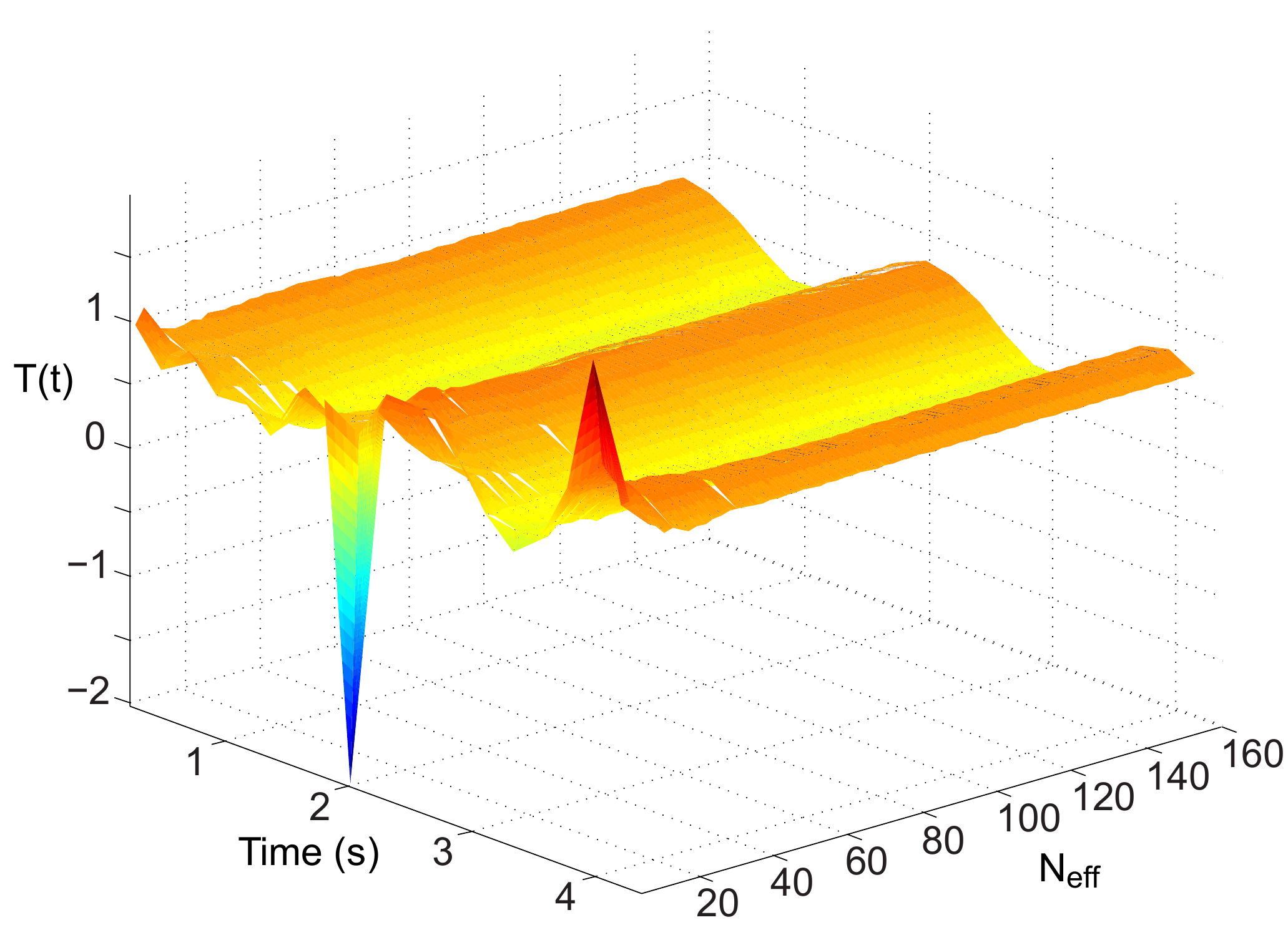}
\caption{Reconstructed sinusoidal signal as a function of the effective number of pixels.}
\label{signalvsNeff}
\end{figure}

In order to measure the $SNR$, several single shot measurements were performed with the same experimental parameters and the WP controller driven by different continuous voltages such as the transmission coefficient is set as constants during the acquisition time of the ghost images ($T=$0.4, 0.9 and 1). The measured values are $0.38\pm 0.03$, $0.90\pm 0.06$ and $0.99\pm 0.08$, respectively. Using Eq. \ref{eq3}, it corresponds to $SNR$s of $26\pm 8$, $28\pm 6$ and $26\pm 8$, which are in rather good agreement with the theoretical $SNR$ : $\frac{153}{\sqrt{19}}=35$. 
While in \cite{ryczkowski_ghost_2016} the accuracy of the reconstructed temporal signal depends on the number of realizations, in our device the accuracy depends on the number of independent pixels used, in each image, to spatially multiplex the temporal signal. To demonstrate how accuracy is degraded when the number of independent effective pixels decreases, we have repeated the calculations with different sizes of ROI. Fig. \ref{signalvsNeff} shows the reconstructed temporal signal as a function of the number of effective pixels. When ROI with few tens of effective pixels are used, the accuracy of the reconstructed signal is very poor while ROI with few thousands of pixels are necessary to retrieve a temporal signal with a good $SNR$. In the supplemental material, Fig. S7 shows that the $SNR$ increases linearly with $N_{eff}$.     

Now let us consider the experiment of multi-spectral temporal ghost imaging. With the same experimental setup and the same protocol, a stack of 10 random binary patterns with three colors (RGB: Red, Green, Blue) is now generated such as the patterns and RGB channels of a pattern are independent. The color patterns are displayed on the LCD screen and images are recorded with the same CMOS camera used in the 24 bits RGB mode. Because the RGB signals are directly encoded in the displayed patterns, the WP and the polarizer are removed for this experiment. During the long exposure time of the camera, color images are modulated with different temporal binary signals that are multiplexed over the RGB channels of the displayed patterns. In order to avoid any crosstalk between the RGB channels, images are saved in a RAW format. Then, the RGB channels of the recorded images are numerically separated and with Eq. \ref{eq2}, the RGB temporal signals are estimated by calculating the intensity correlations for each channel.
 
\begin{figure}[ht]
\centering
\includegraphics[width=8cm]{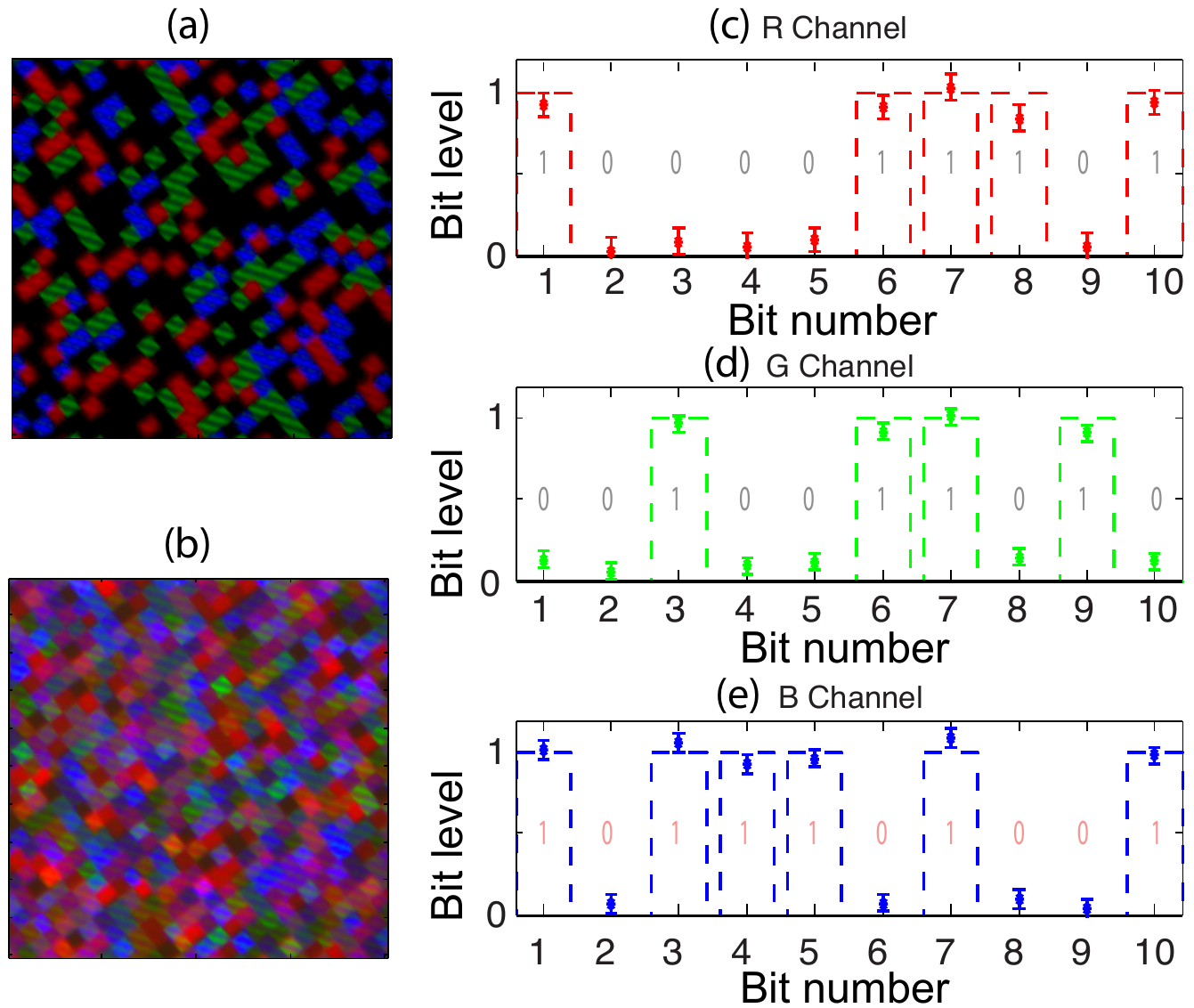}
\caption{Zoom on $700\times 700$ pixels images of a single RGB random pattern (a) and of the time integrated color pattern's stack (b) where wavelength-multiplexed temporal signals are embedded. (c) to (e) single shot measurement of three different binary words of 10 bits multiplexed in the RGB channels of the ghost image. The error bars are deduced from the measured $SNR$. The dotted curves show the original temporal signals.}
\label{signauxRGB}
\end{figure}

Figures \ref{signauxRGB} show typical images of a color pattern of the stack (\ref{signauxRGB}a) and of a ghost image (\ref{signauxRGB}b) where three different temporal signals, encoded on the RGB channels of the patterns, are embedded. Pixels are lightened with a probability $p=0.5$ and the RGB color of a pixel is randomly defined with the same probability. Stack with 10 patterns is generated and the display time of patterns on the LCD screen is 200 $ms$. Figures \ref{signauxRGB}c to \ref{signauxRGB}e show the single shot measurement of three different binary words of 10 bits multiplexed in the RGB channels of the ghost image. The error bars are deduced from the measured $SNR$ and the dotted curves show the original binary words. We can also point out that when only blue patterns are displayed (bits 4 and 5 of the temporal sequence), the correlation levels in the red and the green channels are almost null. This confirms that the crosstalk between channels is negligible.  
 
\section{Conclusion}
To summarize, these experiments represent the first demonstration of ghost imaging of non-reproducible time objects. They are performed by multiplying these objects with computer-generated random monochrome or color images, ensuring the exact space-time transposition of computational ghost imaging and wavelength-multiplexing ghost imaging, respectively. With a very simple device we were able to reconstruct with a very good accuracy different kinds of temporal signals. We demonstrate that our device is also able to separate and reconstruct accurately wavelength-multiplexed temporal signals.
When compared to the recent experimental demonstration of temporal ghost imaging, the main advantage of our system consists with the replacement of thousands synchronized replica of the temporal signal required in \cite{ryczkowski_ghost_2016} by the use of a detector array with thousands pixels (the camera) that has no temporal resolution.
In the present form of the set-up, its obvious drawback is the slowness which was imposed by the simple and low cost devices used to display the random patterns and to generate the temporal signals. More fundamentally, if temporal distortions due to propagation before time integration do not modify the performances, spatial distortions must be smaller than the effective pixel size. The generation rate of random patterns with a large number of independent spatial modes can be easily increased by using either a fast digital micromirror device to display the random patterns up to tens of $kHz$ \cite{gu_digital_2015} or spatial multiplexing of temporally modulated light sources, like 2D VCSELs array \cite{grabherr_vcsel_2014}, through a multimode optical fiber (MOF) or a complex medium to generate random patterns with a rate up to 8 $GHz$. Indeed,the propagation of light in a MOF or in a complex medium produces deterministic speckle patterns (i.e. random patterns) with many spatial modes (i.e. independent pixels) that can be quickly controlled and addressed \cite{defienne_two-photon_2016,andreoli_deterministic_2015}.
More prospectively, with the development and integration of cameras (e.g. $STAMP$ \cite{nakagawa_sequentially_2014}, $CUP$ \cite{gao_single-shot_2014} or $SPAD$ \cite{gariepy_single-photon_2015} cameras) or spatial encoding technologies that can operate at $THz$ frame, our method offers possible perspectives to perform accurate single-shot measurement of any weak, unique and fast spatio-temporal phenomenon that will affect either the light source before transmission through a transparent or complex medium or the transmission of light through these media.
Finally, the number of channels for multi-spectral temporal ghost imaging can be seriously improved by designing a CCD or CMOS array with a more complex Bayer-like filter placed in front of the pixels. 

\section*{Authors contributions}
F. Devaux and E. Lantz wrote the manuscript text. P.A. Moreau had the original idea of the experiment. F. Devaux designed the experimental setup and performed the experiments. E. Lantz developed the theoretical model. S. Denis participated to the experiments.

\section*{Funding}
This work was supported by the Labex ACTION program (ANR-11-LABX-0001-01).

\bibliography{GhostImaging}


\end{document}